\documentclass{PoS}
\usepackage{microtype,graphicx,amsmath,amssymb,wrapfig}
\let\OLDthebibliography\thebibliography
\renewcommand\thebibliography[1]{
  \OLDthebibliography{#1}
  \setlength{\parskip}{0pt}
  \setlength{\itemsep}{0pt plus 0.3ex}}
\title{Hadron tomography in meson-pair production and gravitational form factors}
\ShortTitle{Hadron tomography in meson-pair production and gravitational 
            form factors}
\author{S. Kumano$^{\,\, a,b}$, \speaker{Qin-Tao Song}$^{\, a}$, 
        and O. V. Teryaev$^{\, c}$ \\
$^a$ KEK Theory Center, Institute of Particle and Nuclear Studies, KEK,\\
\ \ \            and Department of Particle and Nuclear Physics,
Graduate University for Advanced Studies \\
\ \ \  (SOKENDAI),       Ooho 1-1, Tsukuba, Ibaraki, 305-0801, Japan\\ 
$^b$ J-PARC Branch, KEK Theory Center,
     Institute of Particle and Nuclear Studies, KEK,\\
\ \ \ and Theory Group, Particle and Nuclear Physics Division, J-PARC Center,\\
\ \ \ 203-1, Shirakata, Tokai, Ibaraki, 319-1106, Japan\\
$^c$ Bogoliubov Laboratory of Theoretical Physics,
            Joint Institute for Nuclear Research,\\ 
\ \ \ 141980 Dubna, Russia}

\abstract{Generalized parton distributions (GPDs) are 3-dimensional (3D) structure functions for hadrons, and they are important for solving 
the proton spin puzzle including partonic orbital-angular-momentum contributions. The $s$-$t$ crossed quantities 
of the GPDs are generalized distribution amplitudes (GDAs). Here, $s$ and $t$ are Mandelstam variables. The GDAs can be studied in two-photon processes ($\gamma^* \gamma\to h \bar h$)
at KEKB. A GDA describes the amplitude from quark and antiquark to the hadron pair $h \bar h$. 
In 2016, the Belle collaboration reported measurements for pion-pair production in electron-positron collision, 
and the pion GDAs were determined in this work by analyzing the Belle data. In our analysis, the pion GDAs are expressed by a few parameters, 
which are determined by analyzing the Belle data. From the obtained GDAs, form factors of energy-momentum tenor, so called gravitational form factors, are calculated for pion 
in the timelike region. The spacelike gravitational form factors are calculated from the timelike ones by using the dispersion relation. 
Then, the mass radius is calculated as 0.32-0.39 fm and the mechanical radius, defined by the slope of the form factor $\Theta_1$, is calculated as 0.82-0.88 fm for the pion by using the spacelike form factors. This is the first study on gravitational form factors and radii of hadrons from actual experimental measurements. In 2019, the Belle II collaboration will start collecting data by
the higher luminosity Super KEKB, so that the GDAs of other hadrons should also be investigated in the near future. Our studies are valuable in understanding 3D structure and gravitational properties of hadrons.}

\FullConference{23rd international spine symposium (Spin 2018)\\
		September 10-14, 2018\\ Ferrara, Italy}

\begin{document}

\section{Introduction}
\vspace{-0.10cm}

In deeply virtual Compton scattering (DVCS), generalized parton 
distributions (GPDs) can be investigated. From the GPDs, 
one can obtain partonic orbital-angular-momentum contributions 
to the nucleon spin. Therefore, the GPDs are key quantities 
to solve the proton spin puzzle. 
The GPDs are one type of 3-dimensional (3D) structure functions, 
and they reveal internal structure for hadrons.
Their forward limit and moments are 
parton distribution functions (PDFs)
and spacelike form factors.
Other 3D structure functions are generalized distribution 
amplitudes (GDAs), which are the $s$-$t$ crossed quantities of 
the GPDs in the Mandelstam variables $s$ and $t$.
The GDAs prove us information on timelike form factors and 
distribution amplitudes (DAs). They are also useful for 
understanding the GPDs by the $s$-$t$ crossing.
 
In order to find the GDAs, one may use the two-photon process 
$\gamma^* \gamma \rightarrow h \bar{h}$ 
\cite{Diehl:1998dk, Diehl:2000uv, Polyakov:1998ze,Kawamura:2013wfa,Kumano:2017lhr} 
and $\gamma^* N \rightarrow h \bar{h} N$ \cite{Anikin:2004ja}.  
In the latter reaction, both GPDs and GDAs are involved. 
However, the amplitude of $\gamma^* \gamma \rightarrow h \bar{h}$ is 
associated only with the GDAs, so that it is possible to find the GDAs
from experimental measurements.
The two-photon process can be measured in the $e^+ e^-$ collision 
at the KEK B factory, and the appropriate kinetics range should be chosen
to satisfy a factorization condition for extracting the GDAs.
In 2016, the Belle collaboration released measurements of differential 
cross sections for $\gamma^* \gamma \rightarrow \pi^0 \pi^0$
\cite{Masuda:2015yoh}.
One can obtain the pion GDAs by analyzing the Belle data 
\cite{Kawamura:2013wfa,Kumano:2017lhr}. Moreover, 
timelike form factors can be calculated from the determined GDAs
\cite{Kumano:2017lhr}.
In 2018, the KEKB facility was updated with a higher luminosity 
of $8 \times10^{35} $cm$^{-2} $s$^{-1}$, and the Belle II collaboration
started taking data. There will be precise data 
of $\gamma^* \gamma \rightarrow h \bar{h}$ for other hadrons, 
so the GDAs of other hadrons can be investigated as well.

\vspace{-0.10cm}
\section{Pion GDAs in the process $\gamma^* \gamma \rightarrow \pi^0 \pi^0$}
\vspace{-0.10cm}

\begin{figure}[b]
\vspace{-0.3cm}
\centering
\includegraphics[width=0.5\textwidth]{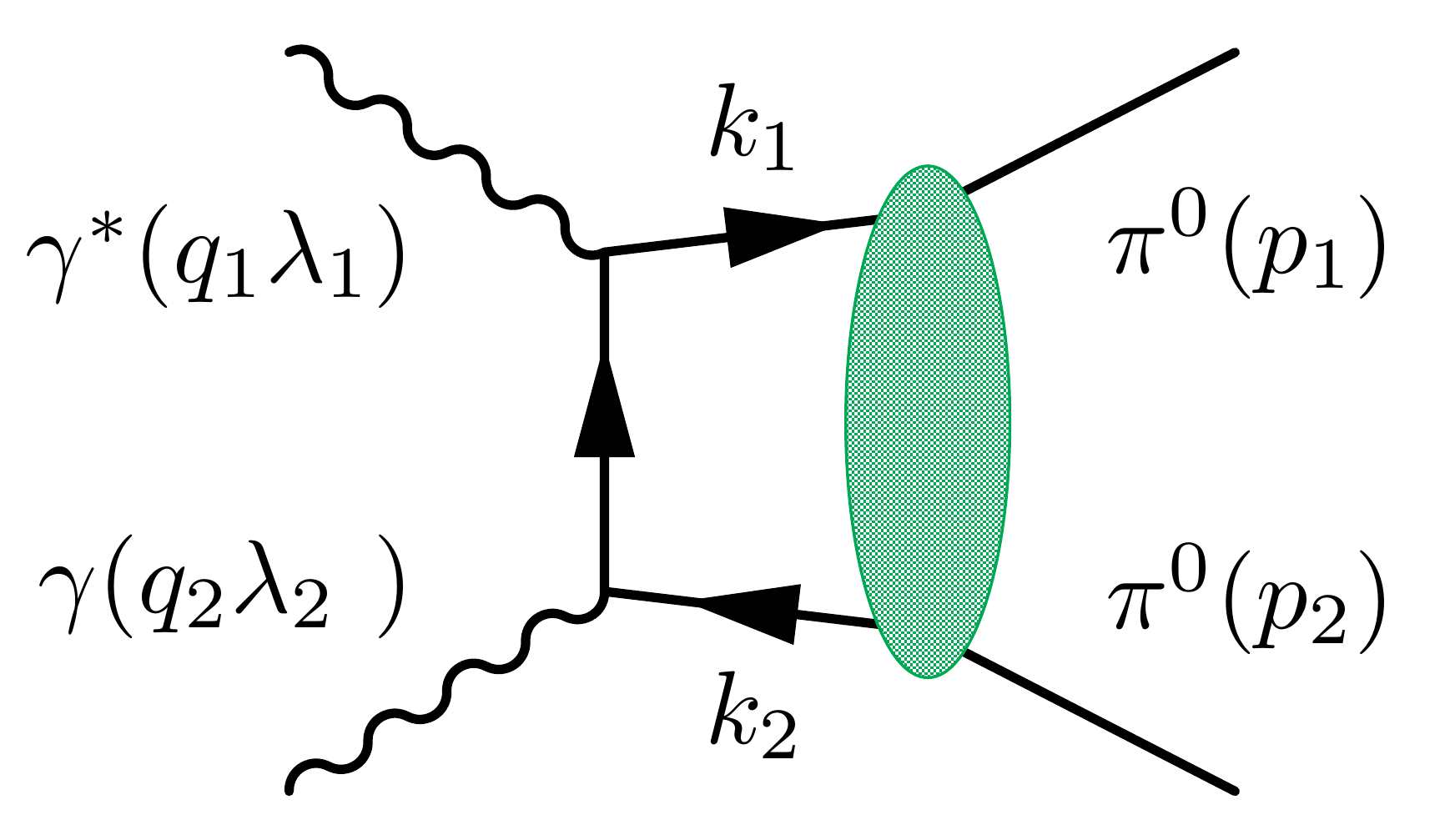}
\vspace{-0.35cm}
\caption{Two-photon process $\gamma^* \gamma \rightarrow \pi^0 \pi^0$.
       The soft part is the pion GDA.}
\label{fig:pion-gda}
\end{figure}
In the DVCS, its amplitude can be factorized into a soft part 
and a hard one, and the soft part of the amplitude is described 
by the GPDs. 
If we consider the $s$-$t$ crossed channel of DVCS, it is 
the two-photon process $\gamma^* \gamma \rightarrow h \bar{h}$ shown 
in Fig.\,\ref{fig:pion-gda}. Similarly, the soft part of 
the two-photon process is described by the GDAs, and $Q^2=-q_1^2$ 
should be much larger than the center-of-mass energy $W^2=(p_1+p_2)^2$ to satisfy 
the factorization condition.
The GDA describes the amplitude of a quark-antiquark pair 
to the hadron-antihadron pair. Here, we take $\pi^0$ 
as for the hadron $h$ and define the GDAs as 
\cite{Diehl:1998dk,Diehl:2000uv, Polyakov:1998ze}
\begin{align} 
& \Phi_q^{\pi^0 \pi^0} (z,\zeta,W^2) 
= \int \frac{d y^-}{2\pi}\, e^{i (2z-1)\, P^+ y^- /2}
  \langle \, \pi^0 (p_1) \, \pi^0 (p_2) \, | \, 
 \overline{q}(-y/2) \gamma^+ q(y/2) 
  \, | \, 0 \, \rangle \Big |_{y^+=\vec y_\perp =0} \, ,
\label{eqn:gda-def}
\end{align}
where $P=p_1+p_2$, $z=k_1^+/P^+$ is the momentum fraction of 
a quark, and $ \xi=p_1^+/P^+$ is the momentum fraction of the pion.
In Fig.\,\ref{fig:pion-gda}, $\lambda_1 $ and $\lambda_2$ are 
the helicities of the virtual photon and the real photon, respectively.
We denote helicity amplitudes as $A_{\lambda_1 \lambda_2}$, and there 
are three independent amplitudes $A_{++}$, $A_{0+}$ and $A_{+-}$
by considering the parity invariance. The leading-twist term $A_{++}$ 
can be expressed by the pion GDA as \cite{Diehl:1998dk,Diehl:2000uv}
\begin{align}
A_{++}=\sum_q \frac{e_q^2}{2} \int^1_0 dz \frac{2z-1}{z(1-z)} 
       \Phi^{\pi^0 \pi^0}_q(z, \xi, W^2).  
\label{eqn:amp2}
\end{align}
The amplitude $A_{0+}$ is a higher-twist term, and it is suppressed
by $1/Q$. The amplitude $A_{+-}$ has an additional running coupling 
constant $\alpha_s(Q^2)$ since the gluon GDA is involved. 
Therefore, both $A_{0+}$ and $A_{+-}$ can be neglected at large $Q^2$,
and the differential cross section is expressed 
only by the amplitude $A_{++}$ for $\gamma^* \gamma \rightarrow \pi^0 \pi^0$ 
\cite{Diehl:1998dk,Diehl:2000uv} as
\begin{align}
d\sigma=\frac{1}{4}  \alpha^2  \pi \frac{\sqrt{1-\frac{4m_\pi^2}{s}} }{ Q^2+s} 
|A_{++}|^2  \sin\theta  d\theta,
\label{eqn:amp2}
\end{align}
where $\alpha$ is the fine structure constant, and
$\theta$ is the scattering angle in the c.m. frame of the final pions.

In the large $Q^2$ limit, the GDAs are independent of $Q^2$ and 
their asymptotic form is given by \cite{Diehl:1998dk,Diehl:2000uv}
\begin{align}
&\sum_q \Phi^{\pi^0\pi^0}_q(z, \xi, W^2)=18n_fz(1-z)(2z-1)
[\tilde{B}_{10}(W)+\tilde{B}_{12}(W)P_2(cos\theta)],  
\nonumber \\
&\tilde{B}_{nl}(W)=\bar{B}_{nl}(W) \exp(i\delta_l), \ \ 
\zeta  = \frac{1+\beta \cos\theta}{2}, \ \ 
\beta=\sqrt{1-\frac{4m_\pi^2}{s}}.
\label{eqn:gda}
\end{align}
In Eq. (\ref{eqn:gda}), there are two terms, $\tilde{B}_{10}(W)$ 
for  S-wave $\pi^0 \pi^0$ and $\tilde{B}_{12}(W)$ for D-wave $\pi^0 \pi^0$.
The $\delta_0$ and $\delta_2$ are the S- and D-wave phase shifts 
\cite{Bydzovsky:2014cda, Nazari:2016wio}, 
respectively, in the $\pi\pi$ elastic scattering below the $KK$ threshold. 
Above the  $KK$ threshold, we introduced 
additional phase 
in the GDA analysis.
\begin{figure}[b]
\vspace{0.2em}
\centering
\includegraphics[width=0.5\textwidth]{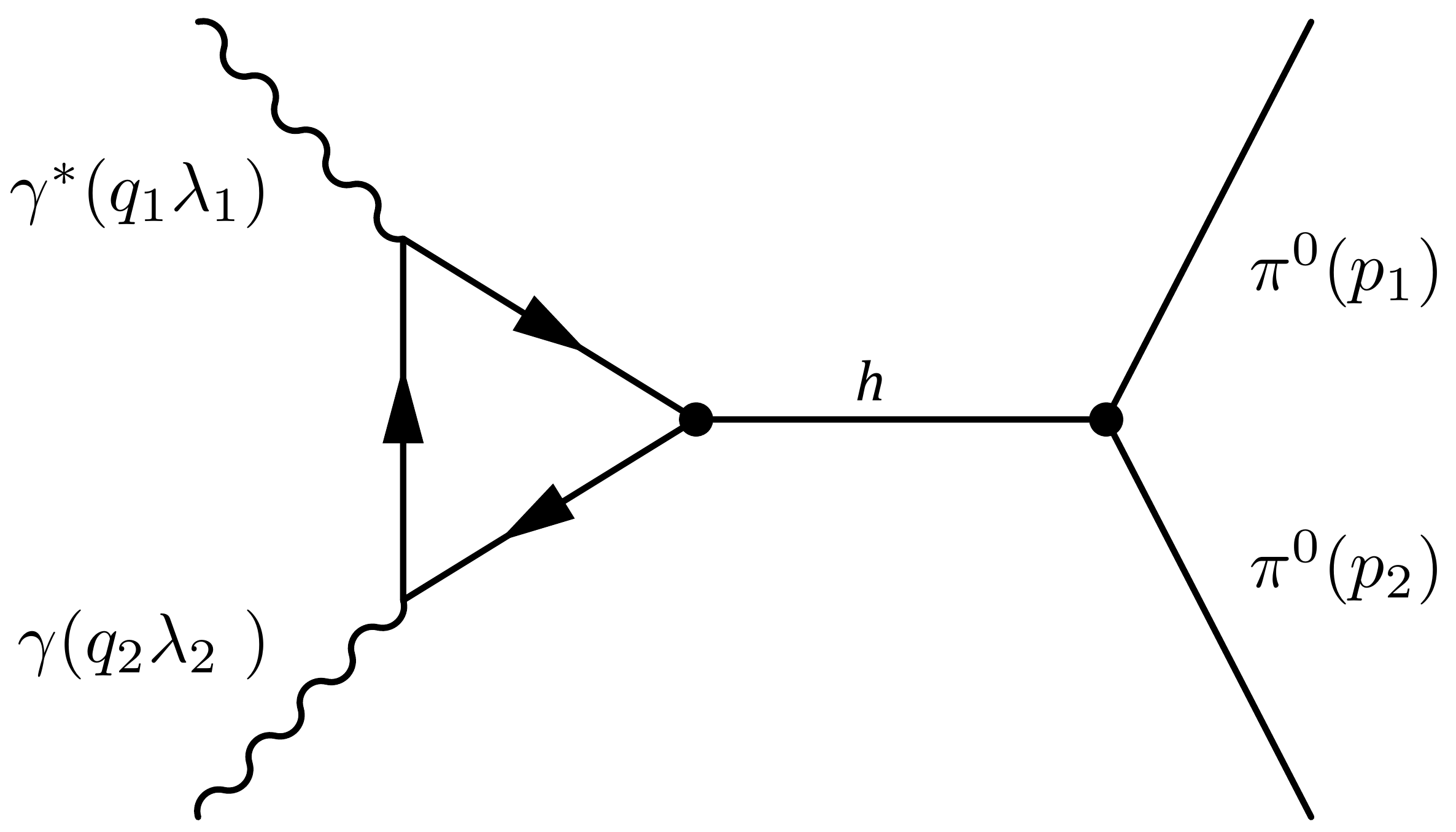}
\vspace{-0.5em}
\caption{Resonance effect through the process 
$\gamma^* \gamma \rightarrow h  \rightarrow \pi^0 \pi^0$.}
\label{fig:resonance}
\end{figure}

In the two-photon process, the resonance effect \cite{Anikin:2004ja} 
is important and it is shown in Fig.\,\ref{fig:resonance}. 
In the GDA analysis, we introduced $f_0(500)$ for an S-wave resonance 
and $f_0(1270)$ for D-wave resonance, whereas $f_0(980)$ is not considered 
in this analysis since it is not clearly shown in the differential 
cross section.

\section{Pion GDA analysis of the Belle data}

In our analysis, the pion GDAs are expressed 
by the addition of continuum terms and resonance ones as
\cite{Kawamura:2013wfa,Kumano:2017lhr} 
\begin{align}
&\Phi_q^{\pi^0 \pi^0}(z, \xi, W^2)=N_h z^\alpha(1-z)^\alpha(2z-1) 
[\tilde{B}_{10}(W)+\tilde{B}_{12}(W)P_2(cos\theta)],  
\nonumber \\
& \tilde{B}_{10}(W)=\left \{ \frac{-3+\beta^2}{2}\frac{10R_{\pi}}{9n_f} F_h(W^2) 
+     \frac{5g_{f_0 \pi\pi} f_{f_0} }{3 \sqrt{2} 
\sqrt{[(M^2_{f_0}-W^2)^2+\Gamma^2_{f_0} M^2_{f_0} ]}}  \right \} 
e^{i\delta_0}     , 
\nonumber \\ 
&  \tilde{B}_{12}(W)=\left \{ \beta^2 \frac{10R_{\pi}}{9n_f} 
F_h(W^2) + \beta^2 \frac{10g_{f_2\pi\pi} f_{f_2} M^2_{f_2} }{9 \sqrt{2}
\sqrt{ (M^2_{f_2}-W^2)^2+\Gamma^2_{f_2} M^2_{f_2} }}  \right \} e^{i\delta_2}.
\label{eqn:gdacb}
\end{align}
Here, the parameter $\alpha$ is predicted as $\alpha=1$ in the large $Q^2$ limit, 
and $R_{\pi}=0.5$ is the momentum fraction carried by quarks in the pion.  
We take $\delta_0$ and $\delta_2$ as the $\pi\pi$ scattering phase shifts 
in the isospin$=$0 channel \cite{Bydzovsky:2014cda, Nazari:2016wio}.
The overall factor $N_h$ depends on $\alpha$ to satisfy the sum rule 
$\int_0^1 dz (2z -1) \, \Phi_q^{\pi^0 \pi^0} (z,\,\zeta,\, 0) 
=-2R_{\pi}\zeta(1-\zeta )$  \cite{Polyakov:1998ze}. 
The resonance effect of $f_0(500)$ appears in the S-wave term 
$\tilde{B}_{10}(W)$, and the D-wave term $\tilde{B}_{12}(W)$ 
contains the resonance effect of $f_2(1270)$. 
The details of Eq. (\ref{eqn:gdacb}) are explained in 
Ref. \cite{Kumano:2017lhr}.

\begin{figure}[hbt]
\vspace{0.2em}
\centering
\includegraphics[width=0.5\textwidth]{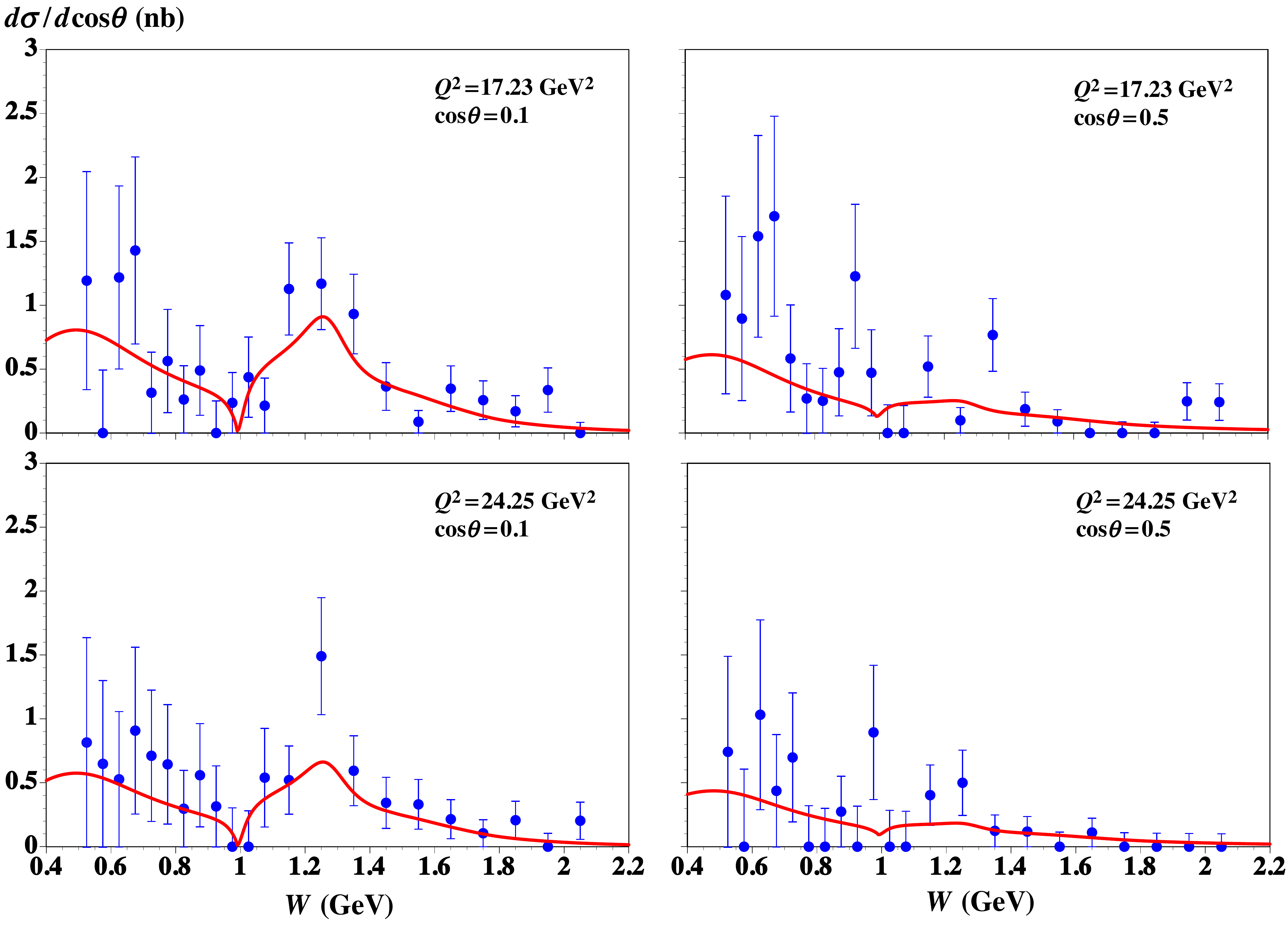}
\vspace{-0.5em}
\caption{$W$ dependence of the cross section 
on  $\gamma^* \gamma \rightarrow \pi^0 \pi^0$
in comparison with Belle measurements. The energy scale $Q^2$ is set as 17.23 GeV$^2$ 
and 24.25 GeV$^2$, and $\cos\theta$ is set as 0.1 and 0.5 
\cite{Kumano:2017lhr}.}
\label{fig:cross}
\end{figure}

There are 5 parameters in our GDA expression,
and they are determined by analyzing the Belle data. 
We show the differential cross section of 
$\gamma^* \gamma \rightarrow \pi^0 \pi^0$ in comparison 
with Belle data in Fig. \ref{fig:cross}, and this GDA analysis 
gives a good description of the experimental data with 
$\chi^2/\text{d.o.f.}=1.09$ \cite{Kumano:2017lhr}. 
In Fig. \ref{fig:cross}, the resonance peak of $f_2(1270)$ 
is seen at $W \simeq 1.2$ GeV, and it is a dominant contribution 
at this invariant-mass region with the Breit-Wigner form 
of the resonance in Eq. (\ref{eqn:gdacb}). 
The resonance effect of $f_0(500)$ is a broad distribution
due to the large decay width.

\begin{figure}[hbt]
\hspace{0.30cm}
\includegraphics[width=0.45\textwidth]{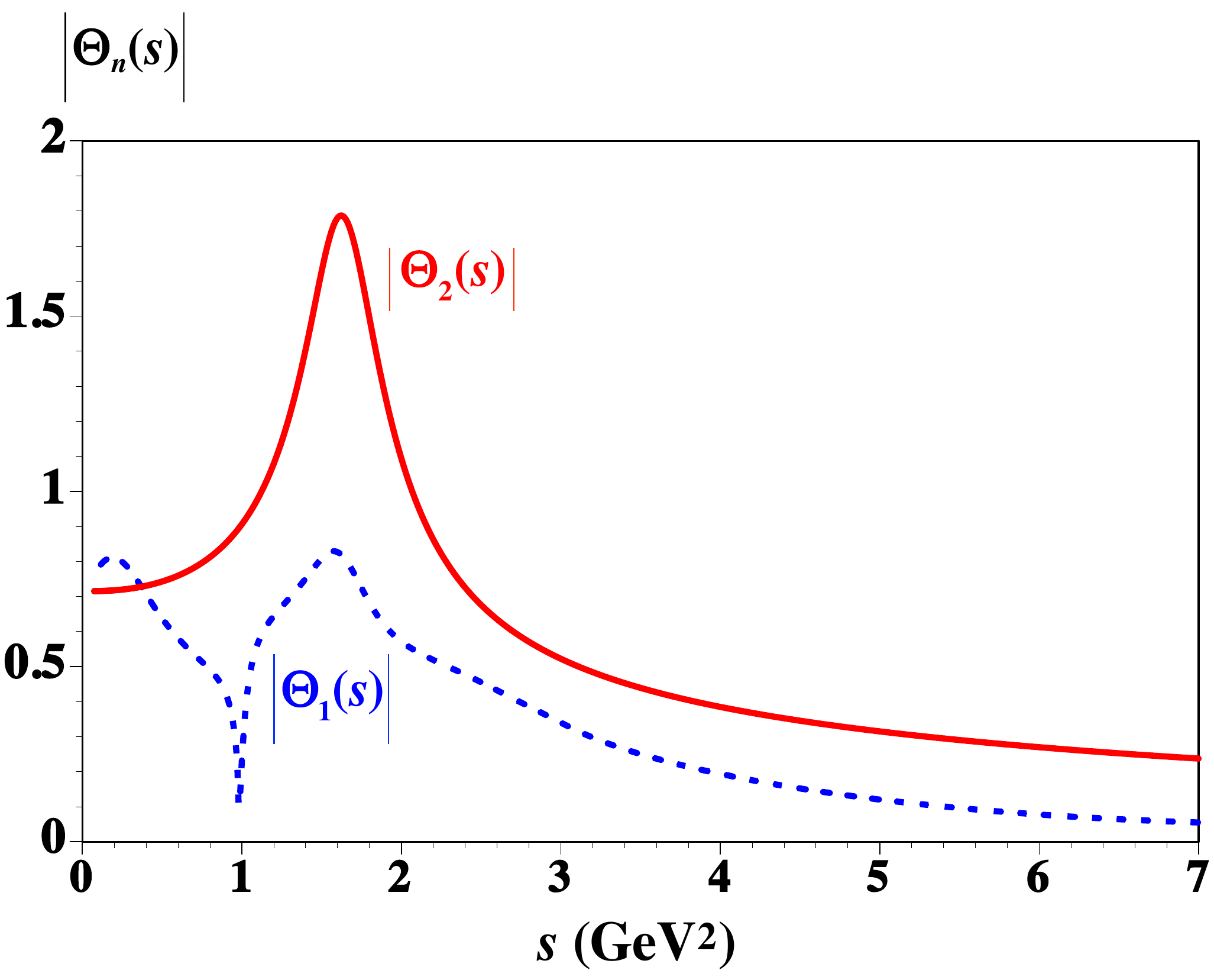}
\hspace{0.30cm}
\includegraphics[width=0.45\textwidth]{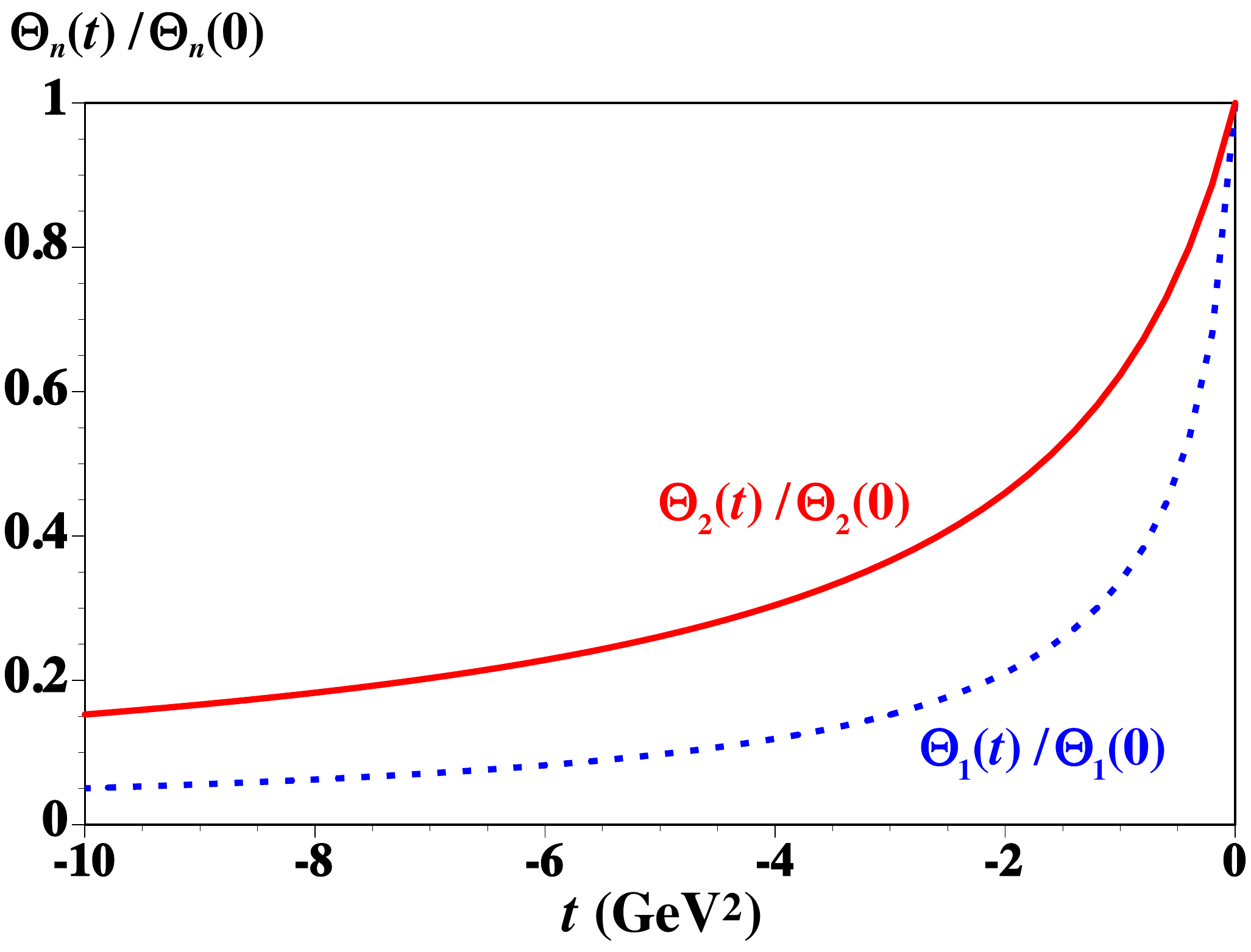}
\hspace{-0.30cm}
\caption{Left panel: The gravitational form factors 
     $\Theta_{1}(s)$ and $\Theta_{2}(s)$ 
     in the timelike region for pion.  
     Right panel: The gravitational form factors $\Theta_{1}(t)$
     and $\Theta_{2}(t)$ in the spacelike region 
     for pion \cite{Kumano:2017lhr}.}
\label{fig:theta}
\end{figure}

The hadronic matrix element of the energy-momentum tensor $T_q^{\mu \nu}$ 
can be expressed by the GPD in the spacelike region. 
Similarly, we can also study the matrix element of 
the energy-momentum tensor for pion in the timelike region 
by the pion GDA, and it is expressed as  \cite{Polyakov:1998ze}
\begin{align}
& \int_0^1 dz (2z -1) \, 
\Phi_q^{\pi^0 \pi^0} (z,\,\zeta,\,W^2) = \frac{2}{(P^+)^2} 
\langle \, \pi^0 (p_1) \, \pi^0 (p_2) \, | \, T_q^{++} (0) \,
       | \, 0 \, \rangle .
\label{eqn:integral-over-z}
\end{align}
The right side of Eq. (\ref{eqn:integral-over-z}) can be expressed 
by two  gravitational form factors $\Theta_{1}$ and  $\Theta_{2}$ as
\begin{align}
& \langle \, \pi^0 (p_1) \, \pi^0 (p_2) \, | \sum_q  \, T_q^{\mu\nu} (0) \, 
| \, 0 \, \rangle 
= \frac{1}{2} 
  \left [ \, \left ( s \, g^{\mu\nu} -P^\mu P^\nu \right ) \, \Theta_{1} (s)
                + \Delta^\mu \Delta^\nu \,  \Theta_{2} (s) \,
  \right ],
\label{eqn:emt-ffs-timelike-0}
\end{align}
where $\Delta=p_1-p_2$. The form factor $\Theta_{1}$ is related 
to the mass or energy, and $\Theta_{2}$ is related to mechanical
properties (pressure and shear force).
The gravitational form factors of the pion can be obtained 
from the determined GDAs \cite{Kumano:2017lhr} as 
 \begin{align}
\Theta_{1} (s) 
= \frac{3}{5} [ \widetilde B_{12} (W^2)-2\widetilde B_{10} (W^2)  ], \ \ 
\Theta_{2} (s)  
= \frac{9}{5 \, \beta^2} \widetilde B_{12} (W^2) .
\label{eqn:emt-ffs-gdas}
\end{align}
In Eq. (\ref{eqn:emt-ffs-gdas}),  $\Theta_{1}(s)$ contains both S-wave 
and D-wave contributions, and $\Theta_{2}(s)$ contains only the D-wave one.
We show the timelike gravitational form factors $\Theta_{1}(s)$ 
and $\Theta_{2}(s)$ in the left panel of Fig.\,\ref {fig:theta}, 
and the resonance effect of $f_2(1270)$ is seen around 
$s=$1.5 GeV$^2$ in both $\Theta_{1}$ and  $\Theta_{2}$.

By the dispersion relation \cite{form-factor-dispersion}, 
the timelike gravitational form factors can be converted 
to the spacelike gravitational form factors:
\begin{align}
F^h (t) & = \int_{4 m_h^2}^\infty \frac{ds}{\pi} 
            \frac{{\rm Im}\, F^h (s)}{s-t-i\varepsilon}.
\label{eqn:dispersion-form-1}
\end{align}
We plot the normalized spacelike gravitational form factors 
$\Theta_{1}(t)$ and $\Theta_{2}(t)$ in the right panel
of Fig.\,\ref{fig:theta}, and we find that $\Theta_{1}(t)$ 
decreases faster than $\Theta_{2}(t)$ as $|t|$ increases.

The root-mean-square (rms) radii can be obtained from
the spacelike gravitational form factors $\Theta_{1}(t)$ 
and $\Theta_{2}(t)$ by calculating the slopes of 
gravitational form factors at $t=0$:
\begin{align}
\! \! \!
\langle \,  r^2 \, \rangle _h
=  \left. \frac{6}{F(t=0)} \, \frac{F ^h (t)}{dt} \right|_{|t| \to 0} 
= \frac{6}{F(t=0)}
\int_{4 m_\pi^2}^\infty ds \, \frac{{\rm Im} \, F^h(s)}{\pi s^2}, \ \ 
F(t=0) =
\int_{4 m_\pi^2}^\infty ds \, \frac{{\rm Im} \, F^h(s)}{\pi s} ,
\label{eqn:3D-radius-2}
\end{align}
where $F^h$ is $\Theta_1$ or $\Theta_2$.
The gravitational radii are calculated as
the mass radius $\sqrt {\langle r^2 \rangle _{\text{mass}}}=0.39$ fm 
from $\Theta_{1}$ and the mechanical radius
$\sqrt {\langle r^2 \rangle _{\text{mech}}}=0.82$ fm 
from $\Theta_{2}$ \cite{Kumano:2017lhr}.

In our analysis,  $\delta_0$ and $\delta_2$ are the $\pi \pi$ 
scattering phase shifts below the $KK$ threshold, and we introduced 
the additional phase for S-wave phase shift above the $KK$
threshold.  However, even if we add the additional phase to D-wave 
phase shift above the threshold, the Belle data are equally-well 
explained. In this second analysis, we obtained the mass radius  
$\sqrt {\langle r^2 \rangle _{\text{mass}}}=0.32$ fm 
and the mechanical radius 
$\sqrt {\langle r^2 \rangle _{\text{mech}}}=0.88$ fm. 
Therefore, the gravitational radii are estimated as
\cite{Kumano:2017lhr}
\begin{align}
\sqrt {\langle r^2 \rangle _{\text{mass}}} 
=  0.32 \sim 0.39 \, \text{fm}, \, 
\sqrt {\langle r^2 \rangle _{\text{mech}}} 
 = 0.82 \sim 0.88 \, \text{fm} ,
\label{eqn:g-radii-pion-range}
\end{align}
in our analyses.
The charge radius of pion is 
$\sqrt {\langle r^2 \rangle _{\text{charge}}}=0.672 \pm 0.008$ fm 
\cite{Patrignani:2016xqp}, so that the mass radius is smaller 
than the charge radius and the mechanical one is slightly larger.
About progress on the gravitational form factors, 
one may also look at recent studies \cite{Burkert:2018bqq}.

\vspace{-0.10cm}
\section{Summary}
\vspace{-0.15cm}

The GPDs are expected to solve the proton spin puzzle, because they 
can reveal partonic orbital-angular-momentum contributions.
Moreover, the hadronic matrix elements of the energy-momentum tensor 
can also be studied by the GPDs, which provide us a good way 
to investigate the gravitational form factors since they cannot
be probed experimentally by direct gravitational interactions.
The GDAs are the $s$-$t$ crossed quantities of GPDs, so that
the GDAs can be also used to study the GPDs and gravitational form factors.
In this work, we analyzed the experimental data to obtain the GDAs 
of pion by using the KEKB measurements on the two-photon process
$\gamma^* \gamma \to \pi^0 \pi^0$.
From the determined GDAs, the gravitational form factors
were calculated. Furthermore, the gravitational radii were 
calculated for the pion, and we obtained the mass radius (0.32-0.39 fm)
and mechanical radius (0.82-0.88 fm). 
In the near future, we expect that the Belle II collaboration
will release much precise measurements for the process 
$\gamma^* \gamma \rightarrow h \bar{h}$ for various hadrons
in addition to the $\pi^0$ pair.
Then, those measurements will help us understand much detail
on the hadron structure through the GDAs.

\vspace{-0.10cm}
\acknowledgments
\vspace{-0.15cm}
\noindent 
Q.-T. S is supported by the MEXT Scholarship for foreign students 
through the Graduate University for Advanced Studies.

\vspace{-0.10cm}


\end{document}